\documentclass{article} %
\usepackage{iclr2025_conference,GEM_workshop_2025,times}

\usepackage{amsmath,amsfonts,bm}

\def\eqref#1{equation~\ref{#1}}

\def\1{\bm{1}}

\DeclareMathAlphabet{\mathsfit}{\encodingdefault}{\sfdefault}{m}{sl}
\SetMathAlphabet{\mathsfit}{bold}{\encodingdefault}{\sfdefault}{bx}{n}

\usepackage{hyperref}
\usepackage{url}

\usepackage[capitalize,noabbrev]{cleveref}
\usepackage{caption}
\usepackage{subcaption}
\usepackage{siunitx}
\usepackage{multirow}
\usepackage{graphicx}
\usepackage{booktabs}
\usepackage{makecell}
\usepackage{listings}
\newcommand\pcref[1]{(\cref{#1})}

\title{
An evaluation of unconditional 3D molecular generation methods 
}

\author{Martin~Buttenschoen, Yael~Ziv, Garrett~M.~Morris, \&  Charlotte~M.~Deane\\
Department of Statistics\\
24-29 St Giles', Oxford OX1 3LB \\
United Kingdom \\
}

\iclrfinalcopy %
\begin{document}

\maketitle

\begin{abstract}
Unconditional molecular generation is a stepping stone for conditional molecular generation, which is important in \emph{de novo} drug design. 
Recent unconditional 3D molecular generation methods report saturated benchmarks, suggesting it is time to re-evaluate our benchmarks and compare the latest models.
We assess five recent high-performing 3D molecular generation methods (EQGAT-diff, FlowMol, GCDM, GeoLDM, and SemlaFlow), in terms of both standard benchmarks and chemical and physical validity. 
Overall, the best method, SemlaFlow, has a success rate of 87\% in generating valid, unique, and novel molecules without post-processing and 92.4\% with post-processing. 
\end{abstract}

\section{Introduction}
\label{sec:introduction}

Generating drug-like molecules and their conformations is a common task in rational drug design. In drug discovery, molecules need to be generated conditionally in order to satisfy desired properties, such as having a particular shape or improving upon a lead compound. Unconditional molecule generation has been a hot topic in machine learning research in recent years and many models have been proposed for this task \citep{anderson2019cormorant, satorras2021, peng2023, song2023equivarianta, vignac2023}. 
The focus on unconditionally generating molecules is because it serves as a stepping stone towards the conditional tasks, as adapting an unconditional model for conditional purposes is a common approach in solving machine learning problems. For example, in image generation unconditional samplers are conditioned on a user's textual input to generate images of particular types only. 
In drug discovery, there are already examples of this paradigm of first building an unconditional model which is then repurposed for conditional generation (also called goal-directed or controllable generation) \citep{hoogeboom2022, baillif2023deep, xu2023geometric, morehead2024, le2024, ziv2024molsnapper}.

In unconditional generation, the aim is to generate a large set of valid, unique, and novel drug-like molecules. Recent molecular generation methods report saturated benchmarks \citep{irwin2024}. 
However, current testing has not included the assessment of the physical and chemical validity of the output molecules.
Two years ago, \citet{baillif2023deep} lamented the missing 3D assessments in the two benchmarks GuacaMol \citep{brown2019} and MOSES \citep{polykovskiy2020}. 
To improve the validation of the molecular conformations, \citet{baillif2023deep} called for, but did not implement, the use of empirical chemical knowledge to assess the validity of bond lengths, bond angles, dihedral angles, and steric clashes.
These geometry-based tools were, however, recently implemented in the related area of docking by the PoseBusters tool and benchmark \citep{buttenschoen2024posebusters} which uses the RDKit's \citep{landrum2024rdkit} Distance Geometry module that is also used in the ETKDG algorithm \citep{riniker2015better} to check empirically informed upper and lower bounds on various molecular geometries, including bond lengths and bond angles. 
Our work is not the first assessment that uses geometry-based measures, for example, \citet{hoogeboom2022} checked generated 3D conformations against typical bond lengths, but these metrics have not been consistently applied in 3D molecular conformation papers and benchmarks. 

Here we assess five recent high-performing 3D molecular generation methods (EQGAT-diff, Flow-Mol, GCDM, GeoLDM, and SemlaFlow) with and without post-processing, in terms of both standard benchmarks and chemical and physical validity. 

\section{Methods}
\label{sec:methods}

\subsection{Models}

Five recent deep generative models for unconditional 3D molecular generation were tested: EQGAT-diff \citep{le2024}, FlowMol \citep{dunn2024a}, GCDM \citep{morehead2024}, GeoLDM \citep{xu2023geometric}, and SemlaFlow \citep{irwin2024}.
Although full details for each method can be found in the original publications, we provide a short description of each:

\emph{GeoLDM} \citep{xu2023geometric} employs a latent diffusion framework and an autoencoder with a continuous latent space. 
The model was designed to specifically capture roto-translational equivariance constraints with the aim of modelling molecular geometries accurately. GeoLDM does not explicitly predict bonds.

\emph{EQGAT-diff} \citep{le2024} leverages E(3)-equivariant diffusion processes that integrate continuous atomic positions with categorical atomic elements and bond types. 
The authors report that their use of time-dependent loss weighting improves training convergence, sample quality, and inference time. 

\emph{Geometry-Complete Diffusion Model} (GCDM) \citep{morehead2024} incorporates geometry-aware graph neural networks into a denoising process in an attempt to capture molecular geometries effectively. 
A reported key feature of GCDM is its ability to account for chirality. 

\emph{FlowMol} \citep{dunn2024a} uses a flow-matching generative modelling framework. In particular, the FlowMol model combines a continuous framework for atomic positions with discrete state spaces for atom types and bonds.
According to the authors, the lightweight design of the model enables high performance on large datasets such as GEOM-Drugs. 

\emph{SemlaFlow} \citep{irwin2024} trains a scalable E(3)-equivariant message-passing model using conditional flow matching. 
The authors claim a novel molecular size-dependent prior that enhances generative performance, and they write that the overall model has up to 2-orders-of-magnitude shorter sampling times compared to other methods.
 
The five models were developed for the `Drugs' subset of the Geometric Ensemble of Molecules (GEOM) dataset curated by \citet{axelrod2022}. EQGAT-diff, FlowMol, and SemlaFlow use the training, validation, and test splits generated by \citet{vignac2023} and GCDM and GeoLDM use the splits generated by \citet{anderson2019cormorant}. The model weights made available by the authors of each model were used.

\subsection{Assessment}

The five methods were benchmarked for their ability to generate 100,000 valid, novel, unique, drug-like molecules. Only molecules that are valid, novel, and unique are counted as successes. 

Validity can be divided into the validity of the molecular graph and that of the molecular conformation. The molecular graph is checked to be chemically valid---fulfils chemical valency rules, and the molecular conformation is checked to be physically valid---has valid bond geometries and low strain energies. Formally, we say that a molecule is \emph{valid} if its molecular graph is \emph{chemically valid} and its conformation is \emph{physically valid}. 

Chemical validity of the molecular graph is assessed using four tests. 
A molecular graph is \emph{chemically valid} if
1) the generated file can be loaded with the \verb|MolFromMolFile| function of the RDKit \citep{landrum2024rdkit} with the sanitization option turned off;
2) the generated RDKit molecule object can be sanitised using the RDKit's \verb|SanitizeMol| function;
3) the molecule has all of its hydrogens added explicitly, assuming that the molecule is not a radical; and
4) the molecule is connected, that is the generated molecular graph is connected in the mathematical sense and in the chemical sense does not have more than one component (or fragment).
These four tests assess whether a molecular graph is chemically valid. 

Physical validity of the molecular conformation is assessed using six tests. 
A molecular conformation is \emph{physically valid} if it passes the bond lengths, bond angles, planar aromatic rings, planar double bonds, internal steric clash, and internal energy tests of the PoseBusters test suite \citep{buttenschoen2024posebusters}.
Bond lengths and bond angles are compared to experimentally determined values and violations below and above 25\% are flagged. 
Internal steric clash is measured using typical van der Waals radii and violations below 30\% are flagged.
Strain energy is measured by the ratio of the observed energy and that of an ensemble of energy-minimised generated conformations. Here, the Universal Force Field \citep{rappe1992uff} was used with a threshold ratio of 100. 
If all of these tests of the geometry of the conformation pass, then the molecular conformation is physically valid.

Uniqueness and novelty of the molecules is assessed using canonical SMILES strings, generated by the RDKit's \verb|MolToSmiles| function \citep{landrum2024rdkit}.
Formally, a generated molecule is \emph{novel} if its SMILES string does not occur in the reference set, and a molecule is \emph{unique} in a multiset of molecules if its SMILES string occurs only once. 
For the novelty check, the reference set is the entire GEOM Drugs set \citep{axelrod2022}, which was used for the training of all of the benchmarked methods. 

Two standard metrics are computed to compare the distributions of the molecules under these metrics. Drug-likeness is estimated using the quantitative estimate of drug-likeness (QED) \citep{bickerton2012quantifying}, and synthetic accessibility (SA) is approximated using the SAscore \citep{ertl2009}. In addition, the appendix contains the distributions of the spacial score \citep{krzyzanowski2023}, Crippen's logP \citep{wildman1999prediction}, the number of heavy atoms, and the conformation strain approximated using the energy ratio from PoseBusters \citep{buttenschoen2024posebusters}. All underlying metrics were calculated using the RDKit. 
The distribution of the metrics are plotted using the \verb|kdeplot| function of the Python package \verb|seaborn| \citep{waskom2021seaborn}.

Furthermore, ECFP4 count fingerprints \citep{morgan1965generation} with 2048 bits are calculated using the RDKit's \verb|GetMorganGenerator| method. The most prevalent substructures in all generated molecules and reference data sets were identified using the Sort and Slice method \citep{dablander2024sort}. The count vectors were projected into two dimensions using the UMAP algorithm \citep{mcinnes2020umap} with the settings \verb|metric="manhattan"|, \verb|min_dist=1.0|, and \verb|spread=1.0| to visualize chemical space.
Furthermore, the Fréchet ChemNet distance \citep{preuer2018frechet} is calculated by generating canonical SMILES for all molecules (without subsampling), filtering out invalid SMILES, and using the \verb|get_fcd| function of the \verb|fcd| package (version 1.2.2).

As baselines, two data sets are being used: the GEOM Drugs set, which contains 301,855 molecules, including all the training data on which the five methods were trained, and the DrugBank set containing 2,066 approved drugs. Details of the two data sets can be found in \cref{sec:datasets} 

\subsection{Post-processing}

The generated molecules were post-processed in three steps. 
First, the largest fragment was picked using the RDKit's \verb|LargestFragmentChooser|.
Second, missing explicit hydrogens were filled in with the assumption that the molecule is not a radical.
Third, the molecular conformation was refined by minimising energy using the Universal Force Field \citep{rappe1992uff} implemented in the RDKit.

\section{Results}
\label{sec:results}

\begin{table}
\caption{
Validity, uniqueness, and novelty of the generated and reference molecules.
The table contains the percentage of successfully generated molecules that are valid, unique, and novel, out of 100,000; for GEOM Drugs and DrugBank the percentages are out of 301,855 and 2,066 respectively. 
Novelty is relative to GEOM Drugs.
All methods generated large numbers of valid, unique, and novel molecules, for example, SemlaFlow generated 87,523 without post-processing and GCDM generated 95,188 with post-processing.
}
\label{tab:valid_unique_novel}
\begin{center}
\begin{tabular}{lS[table-format=3.1]S[table-format=3.1]S[table-format=3.1]}
\toprule
{} & {\unit{\percent}  Valid} & {\unit{\percent}  Valid \&  Unique} & {\unit{\percent}  Valid \& Unique \& Novel} \\
\midrule
EQGAT-diff & 59.7 & 59.7 & 59.5 \\
FlowMol & 59.8 & 59.8 & 59.7 \\
GCDM & 0.2 & 0.2 & 0.2 \\
GeoLDM & 2.9 & 2.9 & 2.9 \\
SemlaFlow & 87.5 & 87.4 & 87.0 \\
\midrule
EQGAT-diff + PP & 84.2 & 84.2 & 84.0 \\
FlowMol + PP & 84.2 & 84.2 & 84.1 \\
GCDM + PP & 95.2 & 95.2 & 95.2 \\
GeoLDM + PP & 69.6 & 69.3 & 69.3 \\
SemlaFlow + PP & 93.1 & 92.9 & 92.4 \\
\midrule
GEOM Drugs & 94.2 & 93.7 & 0.0 \\
DrugBank & 98.8 & 98.2 & 52.6 \\
\bottomrule
\end{tabular}
\end{center}
\end{table}

All five 3D molecular generation methods generated large sets of valid, unique, and novel molecules.
For example, sampling 100,000 times, SemlaFlow generated 87.0\% and FlowMol 59.7\% valid, unique and novel molecules without post-processing. \cref{tab:valid_unique_novel} shows the total share of novel, unique, and valid molecules generated by each method.
Note that GCDM and GeoLDM do not add all hydrogens without post-processing.
For all methods, the limiting factor is validity, as uniqueness and novelty are almost perfect, with more than 99\% of the valid molecules being novel and unique. 
SemlaFlow was also the fastest of the methods \pcref{tab:runtime}, generating 87,523 valid molecules in 3 hours. 

However, the methods are still not as good as the data sets on which they were trained or that they were aiming to capture. 
Both the GEOM Drugs set, which was used for training of the methods, and the DrugBank set, which is a database of known, approved drugs, have almost perfect scores. GEOM Drugs contains 99.8\% chemically valid molecular graphs and 94.2\% valid molecular conformations, while the DrugBank molecules have 100\% valid molecular graphs and 98.8\% valid molecular conformations \pcref{tab:validity_split}. Given that these training data scores are higher than the best model, there is still room for improvement. 

\begin{figure}[!htbp]
\centering

\begin{subfigure}{\textwidth}
\centering
\includegraphics[width=\textwidth]{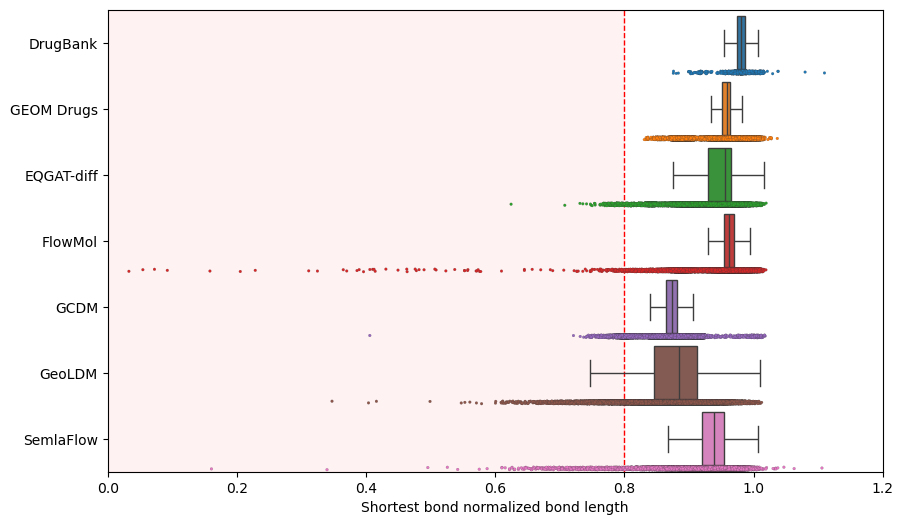} 
\caption{Shortest normalized bond length recorded of each molecule.}
\end{subfigure}

\vspace{10px}

\begin{subfigure}{\textwidth}
\centering
\includegraphics[width=\linewidth]{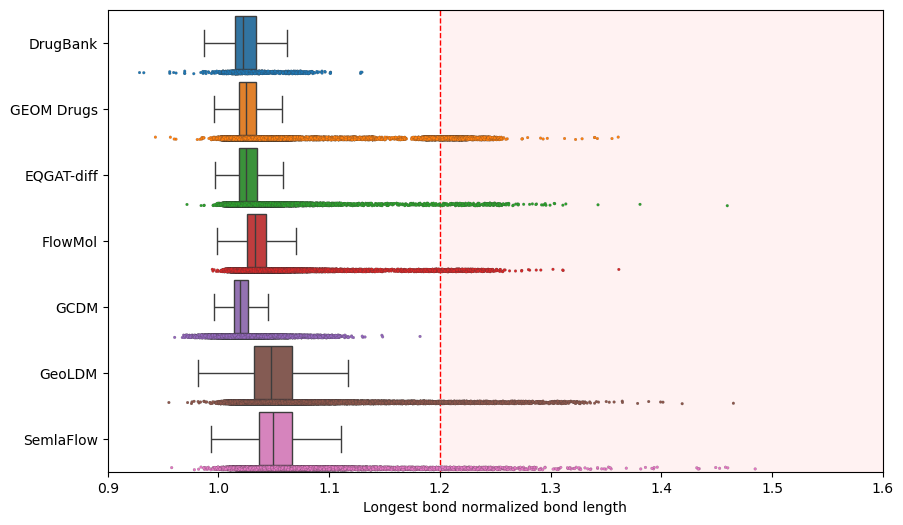} 
\caption{Longest normalized bond length recorded of each molecule.}
\end{subfigure}

\caption{
Distributions of the shortest and longest bond lengths. The values are normalized by the limits obtained from the RDKit's distance geometry module. 
}
\label{fig:bond_lengths}
\end{figure}

\begin{figure}[!htbp]
\centering

\begin{subfigure}{\textwidth}
\centering
\includegraphics[width=\textwidth]{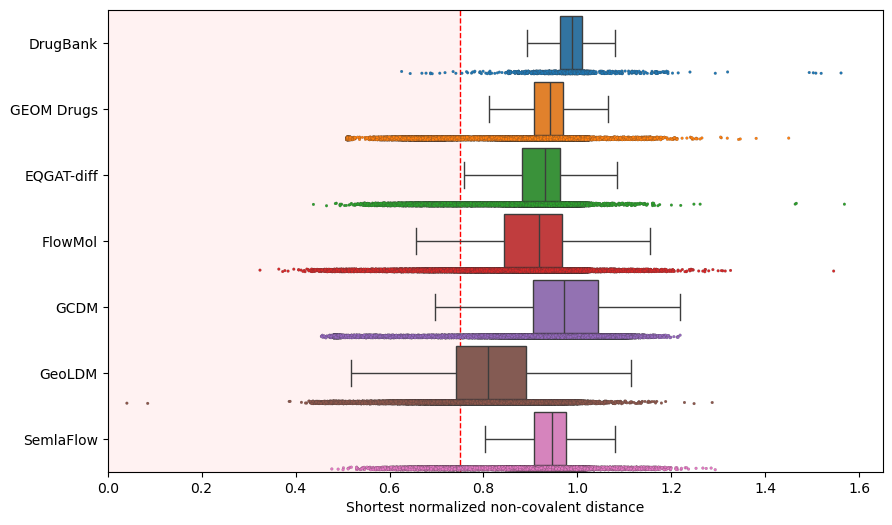} 
\caption{Shortest normalized non-covalent intermolecular distances recorded in each molecule}
\end{subfigure}

\vspace{10px}

\begin{subfigure}{\textwidth}
\centering
\includegraphics[width=\textwidth]{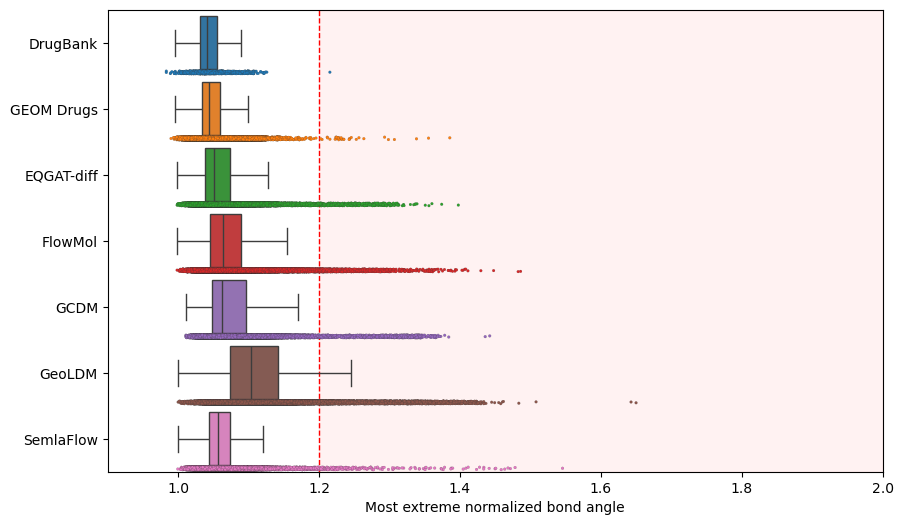} 
\caption{Most extreme normalized bond angle recorded of each molecule.}
\end{subfigure}

\caption{
Distributions of the shortest non-covalent distances and most extreme bond angles. The values are normalized by the limits obtained from the RDKit's distance geometry module. 
}
\label{fig:bond_angles_and_clash}
\end{figure}

The failure modes are in terms of both the generated molecular graphs and the 3D conformations are shown in \cref{tab:validity_split}. For example, EQGAT-diff produces 62.6\% chemically valid molecular graphs and 82.5\% physically valid conformations, leading to an overall validity of 59.7\%. 
For the molecular graphs, the explicit hydrogens check is the largest failure mode for EQGAT-diff, GCDM, and {GeoLDM} \pcref{tab:chemical_validity}, while for the conformational checks, almost all methods show some failures in the geometry-based as well as the energy-based tests \pcref{tab:physical_validity}. Some of the violations that occur are very large (pink zones in \cref{fig:bond_lengths,fig:bond_angles_and_clash}). 
For example, all methods produce some bond lengths that are 20\% too long or too short. 
Despite the fact that the internal steric clash check allows significant overlap of van der Waals radii (30\%), even the best methods---SemlaFlow and GCDM---still exhibited failures.
Overall, there still appears to be potential to improve validity in terms of connectedness and geometry. 

Running the post-processing steps described in the Methods mitigates these results to some degree.
Picking the largest fragment, adding missing hydrogens, and minimising the conformations' energy improved the results by up to 95\% depending on the method. 
For example, SemlaFlow's performance increased by 5.4 percentage points to 92.4\% and GCDM's increased to 95.2\%.
The improvements of GCDM and GeoLDM are mostly due to the addition of hydrogens since these two methods do not generate all hydrogens explicitly.
With post-processing, the best method according to these metrics becomes GCDM. 
In general, post-processing improves the ability of all methods to generate a large number of valid, unique, and novel molecules. 

\begin{figure}[hb]
\centering

\begin{subfigure}{0.95\textwidth}
\centering
\includegraphics[width=\textwidth]{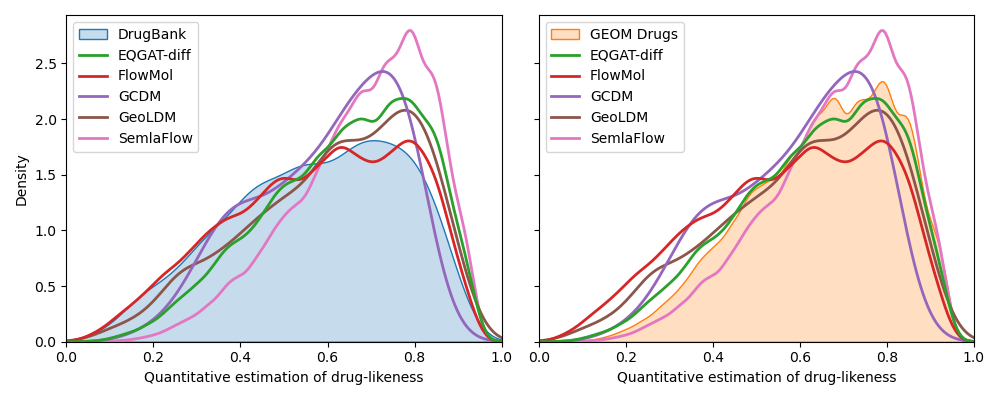} 
\caption{Distributions of the molecules' QED. The QED is a proxy for drug-likeness. The left panel shows the DrugBank molecules in blue as a reference; the right panel shows the GEOM Drugs molecules in light orange as a reference.}
\end{subfigure}

\begin{subfigure}{0.95\textwidth}
\centering
\includegraphics[width=\textwidth]{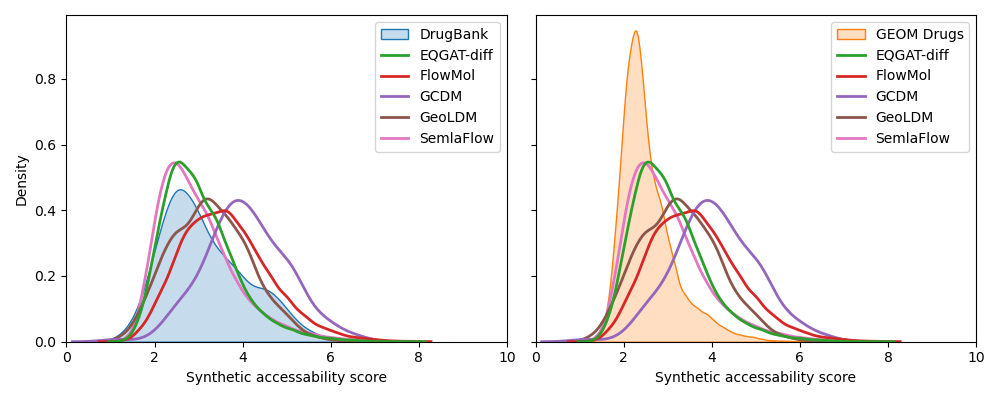} 
\caption{Distributions of the molecules' SAscore. The SAscore is a proxy for synthetic accessibility. The left panel shows the DrugBank molecules in blue as a reference; the right panel shows the GEOM Drugs molecules in light orange as a reference.}
\end{subfigure}

\caption{
Distributions of the molecules in terms of drug-likeness estimated by QED and synthetic accessibility estimated by SAscore in comparison to the approved drugs in DrugBank and to the training data GEOM Drugs.
All methods generate molecules that approximately sit in the same distribution under QED and SAscore as the approved drugs and the training data.
}
\label{fig:qed}
\end{figure}

The distributions of the metrics for drug-likeness and synthetic accessibility of the generated molecules tend to those of the GEOM Drugs training data. \cref{fig:qed} shows the distributions of the QED and SAscore, which estimate drug-likeness and synthetic accessibility respectively. These plots and the additional metrics in the Appendix in \cref{tab:properties} and \cref{fig:properties_1,fig:properties_2} show that the best methods generate molecules that adopt the same distribution under the selected measures as the training data tend to. 

\begin{figure}
\centering

\includegraphics[width=\textwidth,trim={0 0 0 0},clip]{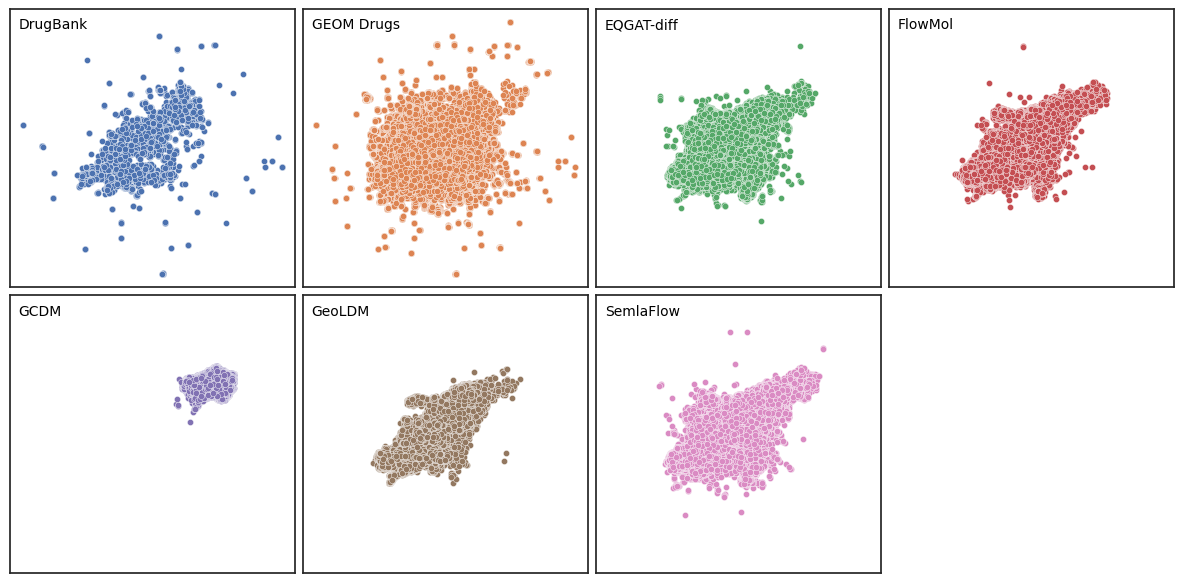} 

\caption{
Visualization of the chemical space covered by the generated molecules and the reference data sets shown using the same projection for all methods. The two-dimensional map was generated from all molecules' ECFP4 2048 count fingerprints by the UMAP algorithm.
}
\label{fig:umap}
\end{figure}

\begin{figure}
\centering
\includegraphics[width=0.55\textwidth]{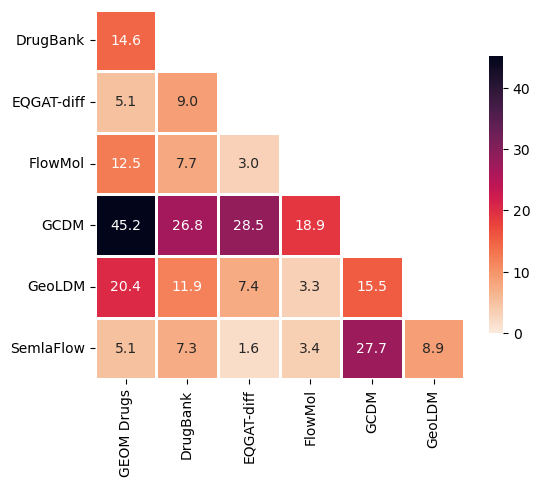} 
\caption{Fréchet ChemNet distance between the sets of generated molecules with post-processing, the training data GEOM Drugs and the approved drugs in DrugBank. SemlaFlow and EQGAT-diff capture the training data the best as they have the lowest distance (5.1) to GEOM Drugs. 
Note that all sets are significantly larger than the recommended data set size (5'000) to calculate this metric except DrugBank which contains 2,066 molecules.}
\label{fig:fcd}
\end{figure}

In terms of chemical diversity, the generated molecules of all methods tend to sit inside the molecules of the training data, but not all methods cover the space sufficiently.
The UMAP projections of the molecules' fingerprints \pcref{fig:umap} show that EQGAT-diff, FlowMol, GeoLDM, and SemlaFlow cover the core of the space of GEOM Drugs but there are also outlier islands on the fringes of the DrugBank and GEOM Drugs distributions, for which the methods do not generate molecules. Also note that GCDM covers a smaller space than the other tested methods. 
These qualitative observations are reflected in the quantitative measurements. The calculated Fréchet ChemNet distance \pcref{fig:fcd} from the GEOM Drugs training data is the smallest for the molecules generated by EQGAT-Diff (5.1) and SemlaFlow (5.1), while GCDM has the largest distance (45.2). 
In terms of this metric, SemlaFlow and EQGAT-diff capture the space of the training data the best. 

\section{Conclusion}
The high observed success rates (of up to 95.2\%) for generating valid, unique, and novel molecules show that these methods are already useful unconditional 3D molecular generators. 
The top methods are able to generate molecules with the same distribution as the training set in terms of the QED and SAscore metrics, but they appear to not fully explore the space of the training data set in terms of ECFP4-2048 count fingerprints and Fréchet ChemNet distance. 
Furthermore, the PoseBusters-based checks used here are still quite generous, for example allowing 30\% closer overlap of the van der Waals radii of atoms. 
Future work should explore further improving these methods against even more stringent physical and chemical checks.

\clearpage
\bibliography{iclr2025_conference}
\bibliographystyle{iclr2025_conference}

\clearpage
\appendix

\section{Further results}

\clearpage

\begin{table}
\caption{
Runtime (in hours) for generating 100,000 samples on a Fedora Linux server using 14 CPU cores, 100 GB RAM, and one Nvidia A100 80GB GPU.
The table reports the total time required for each model to complete the generation process which is described in \cref{sec:method_details}.
}
\label{tab:runtime}
\begin{center}
\begin{tabular}{lS[table-format=3.0]}
\toprule
{} & {Runtime (h)}  \\
\midrule
EQGAT-diff  & 123  \\
FlowMol     & 6  \\
GCDM        & 42  \\
GeoLDM      & 176  \\
SemlaFlow   & 3 \\
\bottomrule
\end{tabular}
\end{center}
\end{table}

\begin{table}
\caption{
Validity of the generated and ground truth molecules.
The table contains the total number of molecules that was to be generated or is contained in the data set and the proportions of molecules that pass all the chemical validity test, all physical validity tests.
The last columns is the percentage of molecules that pass all the tests together.
}
\label{tab:validity_split}
\begin{center}
\begin{tabular}{lS[table-format=3.1]S[table-format=3.1]|S[table-format=3.1]}
\toprule
{} & {\unit{\percent} Chemical} & {\unit{\percent} Physical} & {\unit{\percent} Valid} \\
\midrule
EQGAT-diff & 62.6 & 82.5 & 59.7 \\
FlowMol & 66.4 & 76.2 & 59.8 \\
GCDM & 0.2 & 82.3 & 0.2 \\
GeoLDM & 3.2 & 57.0 & 2.9 \\
SemlaFlow & 91.3 & 90.8 & 87.5 \\
\midrule
EQGAT-diff + PP & 87.1 & 84.2 & 84.2 \\
FlowMol + PP & 87.2 & 84.2 & 84.2 \\
GCDM + PP & 95.5 & 95.2 & 95.2 \\
GeoLDM + PP & 73.8 & 69.6 & 69.6 \\
SemlaFlow + PP & 94.9 & 93.1 & 93.1 \\
\midrule
GEOM Drugs & 99.8 & 94.2 & 94.2 \\
DrugBank & 100.0 & 98.8 & 98.8 \\
\bottomrule
\end{tabular}
\end{center}
\end{table}

\begin{table}
\caption{
Components of the chemical validity of the generated and ground truth molecules. The numbers shown are the
percentages of the molecules that pass each of the tests. The last column is the percentage of
molecules that pass all of the tests.
The chemical tests check the molecular graph generated and they do not check the molecules' 3D conformations.
}
\label{tab:chemical_validity}
\centering
\begin{tabular}{lS[table-format=3.1]S[table-format=3.1]S[table-format=3.1]|S[table-format=3.1]}
\toprule
{} & {Sanitizes} & {Hydrogens explicit} & {Connected} & {\unit{\percent} Chemical} \\
\midrule
EQGAT-diff & 87.1 & 64.4 & 84.4 & 62.6 \\
FlowMol & 86.6 & 81.1 & 68.6 & 66.4 \\
GCDM & 97.2 & 0.2 & 86.6 & 0.2 \\
GeoLDM & 72.5 & 5.3 & 37.4 & 3.2 \\
SemlaFlow & 94.9 & 93.8 & 92.3 & 91.3 \\
\midrule
EQGAT-diff + PP & 87.1 & 87.1 & 87.1 & 87.1 \\
FlowMol + PP & 87.2 & 87.2 & 87.2 & 87.2 \\
GCDM + PP & 95.5 & 95.5 & 95.5 & 95.5 \\
GeoLDM + PP & 73.8 & 73.8 & 73.8 & 73.8 \\
SemlaFlow + PP & 94.9 & 94.9 & 94.9 & 94.9 \\
\midrule
GEOM Drugs & 100.0 & 100.0 & 99.8 & 99.8 \\
DrugBank & 100.0 & 100.0 & 100.0 & 100.0 \\
\bottomrule
\end{tabular}
\end{table}

\begin{table}[!htbp]
\caption{
Components of the physical validity of the generated and ground truth molecules.
The numbers shown are the percentages of the molecules that pass each of the intramolecular PoseBusters tests.
The last column is the percentage of molecules that pass all of the intramolecular tests.
The physical tests check the 3D conformations of the generated molecules.
}
\label{tab:physical_validity}
\centering
\begin{tabular}{lS[table-format=3.1]S[table-format=3.1]S[table-format=3.1]S[table-format=3.1]S[table-format=3.1]S[table-format=3.1]|S[table-format=3.1]}
\toprule
{} & {}        & {}        & {Internal}        & {Planar}   & {Planar} & { }  & {}      \\
{}        & {Bond}    & {Bond}   & {steric}   & {aromatic} & {double} & {Internal} & { } \\
{}        & {lengths} & {angles} & {clash}    & {rings}    & {bonds}  & {energy}   & {\unit{\percent} Physical} \\
\midrule
EQGAT-diff & 87.0 & 86.9 & 82.9 & 87.0 & 87.0 & 86.8 & 82.5 \\
FlowMol & 86.5 & 86.1 & 82.5 & 86.5 & 81.2 & 85.4 & 76.2 \\
GCDM & 97.2 & 96.4 & 94.8 & 97.2 & 97.2 & 84.2 & 82.3 \\
GeoLDM & 71.0 & 69.6 & 62.7 & 72.4 & 71.2 & 70.0 & 57.0 \\
SemlaFlow & 94.6 & 94.8 & 92.0 & 94.9 & 94.2 & 94.8 & 90.8 \\
\midrule
EQGAT-diff + PP & 87.0 & 87.1 & 84.6 & 87.0 & 87.0 & 86.7 & 84.2 \\
FlowMol + PP & 87.2 & 87.2 & 84.9 & 87.2 & 87.1 & 86.8 & 84.2 \\
GCDM + PP & 95.5 & 95.5 & 95.5 & 95.5 & 95.5 & 95.2 & 95.2 \\
GeoLDM + PP & 73.7 & 73.8 & 70.5 & 73.8 & 73.6 & 73.1 & 69.6 \\
SemlaFlow + PP & 94.9 & 94.9 & 93.2 & 94.9 & 94.8 & 94.8 & 93.1 \\
\midrule
GEOM Drugs & 100.0 & 100.0 & 94.6 & 100.0 & 99.7 & 100.0 & 94.2 \\
DrugBank & 100.0 & 100.0 & 99.7 & 100.0 & 100.0 & 99.1 & 98.8 \\
\bottomrule
\end{tabular}
\end{table}

\begin{table}
\caption{
Various molecular properties. The table contains the median and interquartile range of the SAscore,  QED, Lipinski rule of five, the spacial score, molecular weight, number of heavy atoms, number of rings, and logP.
}
\label{tab:properties}
\centering 
\sisetup{separate-uncertainty}

\begin{tabular}{lS[table-format=1.2(1.2)]S[table-format=1.2(1.2)]S[table-format=1.1(1.1)]S[table-format=2.1(2.1)]}
\toprule
{} & {SAscore} & {QED} & {Lipinski R5} & {Spacial score} \\
\midrule
EQGAT-diff & 3.14(1.18) & 0.65(0.27) & 5.0(0.0) & 15.33(8.59) \\
FlowMol & 3.66(1.40) & 0.57(0.34) & 5.0(0.0) & 17.65(12.04) \\
GCDM & 4.99(0.87) & 0.38(0.29) & 5.0(1.0) & 36.84(10.55) \\
GeoLDM & 4.65(1.32) & 0.50(0.35) & 5.0(0.0) & 19.33(8.82) \\
SemlaFlow & 2.82(1.11) & 0.70(0.22) & 5.0(0.0) & 15.35(9.33) \\
\midrule
EQGAT-diff + PP & 3.11(1.22) & 0.65(0.28) & 5.0(0.0) & 15.68(9.31) \\
FlowMol + PP & 3.61(1.43) & 0.59(0.31) & 5.0(0.0) & 17.81(12.81) \\
GCDM + PP & 4.96(0.87) & 0.39(0.29) & 5.0(1.0) & 37.05(10.61) \\
GeoLDM + PP & 4.32(1.39) & 0.49(0.33) & 5.0(0.0) & 21.88(11.86) \\
SemlaFlow + PP & 2.81(1.11) & 0.70(0.22) & 5.0(0.0) & 15.38(9.40) \\
\midrule
GEOM Drugs & 2.39(0.69) & 0.67(0.26) & 5.0(0.0) & 13.74(5.90) \\
DrugBank & 2.96(1.39) & 0.59(0.31) & 5.0(1.0) & 15.69(11.80) \\
\bottomrule
\end{tabular}

\begin{center}
\sisetup{separate-uncertainty}
\begin{tabular}{lS[table-format=3.2(3.2)]S[table-format=2.2(2.2)]S[table-format=1.2(1.2)]S[table-format=1.2(1.2)]}
\toprule
{} & {Weight [Da]} & {\# Heavy atoms} & {\# Rings} & {LogP} \\
\midrule
EQGAT-diff      & 352.16(104.91) & 25.00(8.00) & 3.00(2.00) & 2.55(1.86) \\
FlowMol         & 337.10(98.07) & 24.00(6.00) & 3.00(1.00) & 1.52(2.14) \\
GCDM            & 354.25(58.02) & 24.00(4.00) & 2.00(1.00) & 0.57(2.62) \\
GeoLDM          & 366.19(106.01) & 26.00(8.00) & 3.00(2.00) & 1.08(2.42) \\
SemlaFlow       & 321.05(98.01) & 23.00(7.00) & 3.00(1.00) & 2.77(1.74) \\
\midrule
EQGAT-diff + PP & 349.10(105.00) & 25.00(7.00) & 3.00(2.00) & 2.52(1.88) \\
FlowMol + PP    & 316.15(106.02) & 22.00(7.00) & 2.00(1.00) & 1.42(2.10) \\
GCDM + PP       & 348.24(61.98) & 24.00(4.00) & 2.00(1.00) & 0.56(2.59) \\
GeoLDM + PP     & 308.03(144.10) & 21.00(10.00) & 2.00(2.00) & 0.69(2.34) \\
SemlaFlow + PP  & 317.19(98.08) & 23.00(7.00) & 3.00(1.00) & 2.74(1.75) \\
\midrule
GEOM Drugs      & 351.23(105.00) & 25.00(7.00) & 3.00(2.00) & 2.92(1.64) \\
DrugBank        & 324.13(171.09) & 23.00(12.00) & 3.00(3.00) & 2.58(2.88) \\
\bottomrule
\end{tabular}
\end{center}
\end{table}

\clearpage

\begin{figure}
\centering

\begin{subfigure}{\textwidth}
\centering
\includegraphics[width=\textwidth]{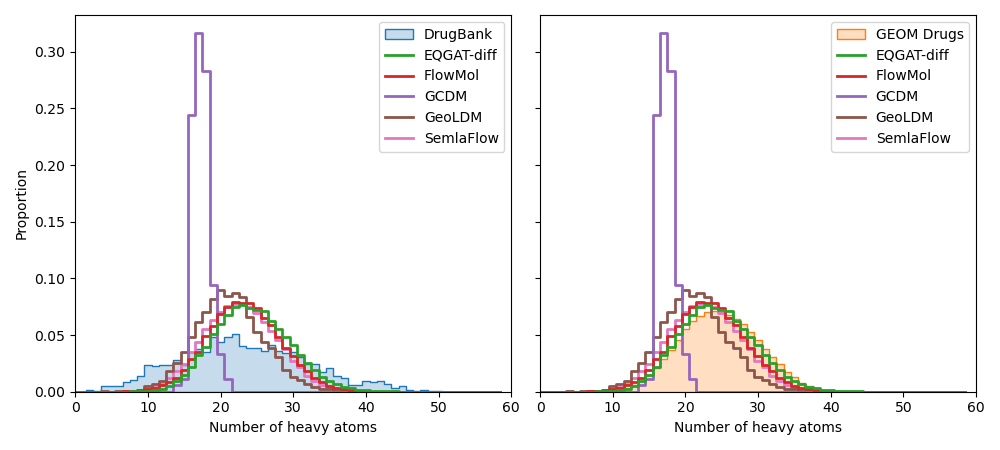} 
\caption{Number of heavy atoms.}
\end{subfigure}

\begin{subfigure}{\textwidth}
\centering
\includegraphics[width=\textwidth]{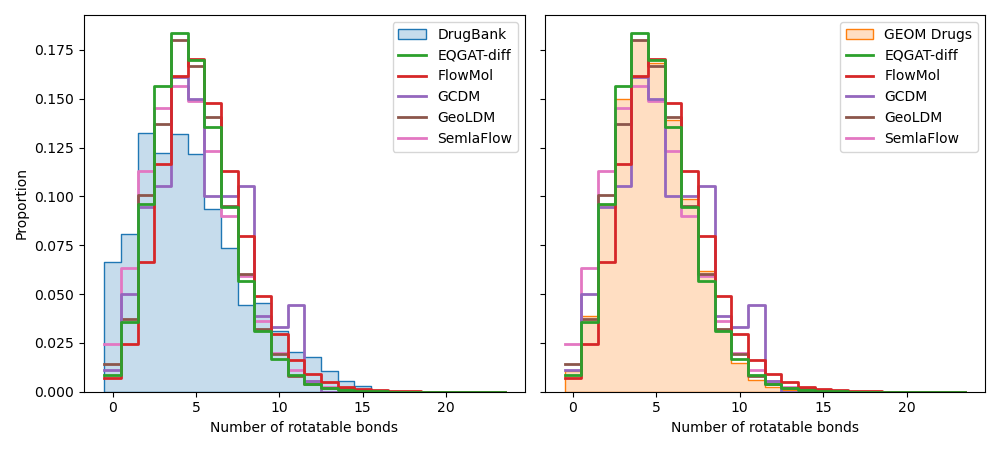} 
\caption{Number of rotatable bonds.}
\end{subfigure}

\begin{subfigure}{\textwidth}
\centering
\includegraphics[width=\textwidth]{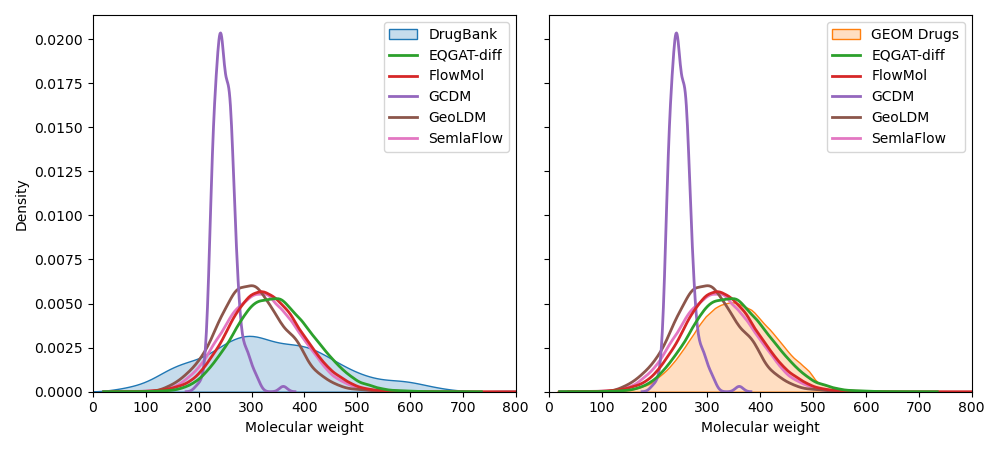} 
\caption{Molecular weight in daltons.}
\end{subfigure}

\caption{
Distributions of the valid molecules in terms of number of heavy atoms, number of rotatable bond, and molecular weight. 
}
\label{fig:properties_1}
\end{figure}

\begin{figure}
\centering

\begin{subfigure}{\textwidth}
\centering
\includegraphics[width=\textwidth]{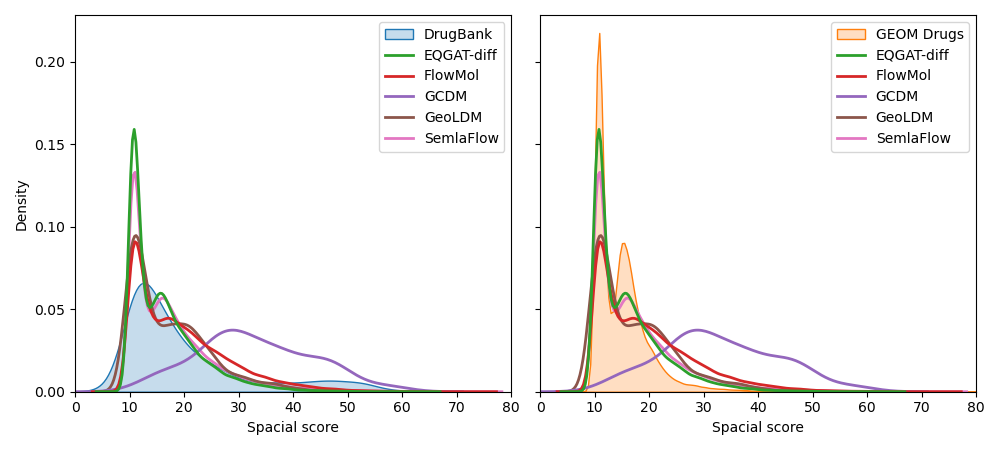} 
\caption{Molecule complexity estimated by the Spacial Score.}
\end{subfigure}

\begin{subfigure}{\textwidth}
\centering
\includegraphics[width=\textwidth]{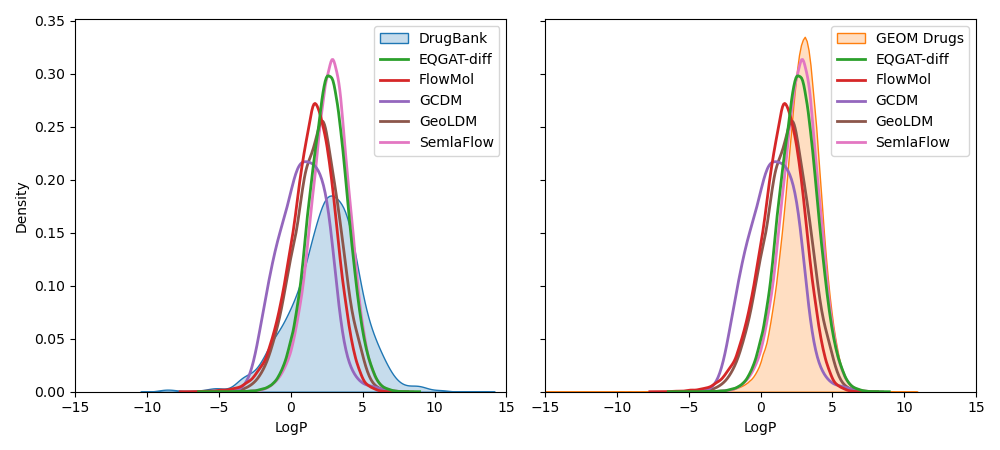} 
\caption{Lipophilicity estimated by Crippen’s logP.}
\end{subfigure}

\begin{subfigure}{\textwidth}
\centering
\includegraphics[width=\textwidth]{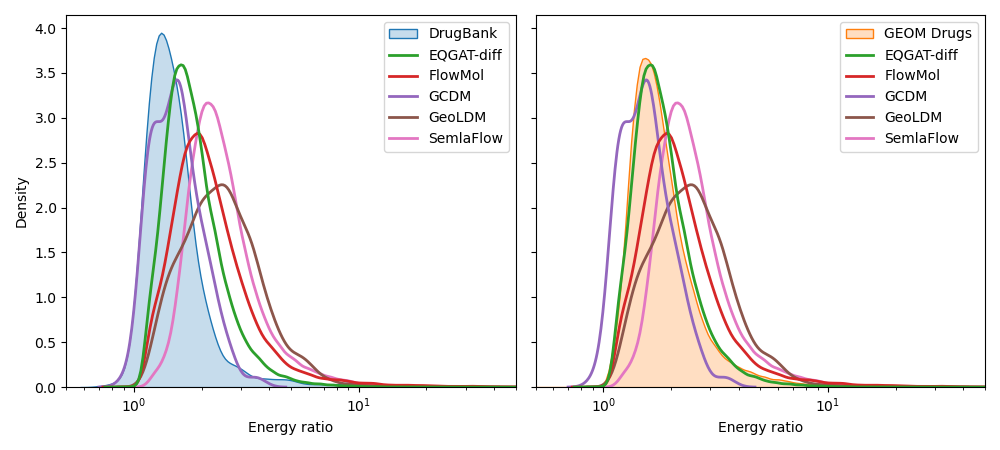} 
\caption{Conformational strain estimated by the energy ratio of the generated conformation relative to the average energy of an ensemble of energy minimization conformations generated with ETKDGv3.
The energies were estimated using the Universal Force Field. 
}
\end{subfigure}

\caption{
Distributions of the valid molecules in terms of synthetic accessibility, molecule complexity, and lipophilicity. 
}
\label{fig:properties_2}
\end{figure}

\clearpage

\section{Details on sampling methods}
\label{sec:method_details}
This section describes how each model was used to generate 100,000 molecules.

\subsection{EQGAT-diff}
The code for EQGAT-diff with commit hash \verb|68aea80691a8ba82e00816c82875347cbda2c2e5| was obtained from the public code repository of the authors \url{https://github.com/tuanle618/eqgat-diff/tree/main}.
The model weights trained on the GEOM Drugs were obtained from the authors upon request.The model was run using the script \verb|run_evaluation.py| with the setting for batch size at 100 and the dataset option to `drugs'.
All other parameters were kept at their default settings. The model was run 20 times sampling 5,000 molecules each time. The generated XYZ files were converted to SDF format and combined into a single file with Open Babel \citep{oboyle2011}.

\subsection{FlowMol}
The code for FlowMol with commit hash \verb|c3503939ce409a12e558e3231b5c807f86d9fe1d| was obtained from the authors’ public code repository \url{https://github.com/Dunni3/FlowMol}. The model weights were obtained from \url{https://bits.csb.pitt.edu/files/FlowMol/trained_models/} and placed in the directory as instructed by the README provided by the authors. 
The script \verb|test.py| was run using \verb|n_timesteps=250|, \verb|nummols=100000|, and \verb|model_dir=flowmol/trained_models/geom_ctmc| and the output was a single SDF file. 

\subsection{GCDM}
The code for GCDM with commit hash \verb|109d9d7625a00fb669454246fc846f348be3df0d| was obtained from the authors' public code repository \url{https://github.com/BioinfoMachineLearning/bio-diffusion}. 
The model checkpoints used were obtained from Zenodo \url{https://zenodo.org/record/13375913/files/GCDM_Checkpoints.tar.gz}. The model was run using script \verb|mol_gen_sample.py|. 
The model was run 50 times using the seeds 123 through 173 generating 2,000 molecules each time. All other settings passed to the script were \verb|num_nodes=44|, \verb|all_frags=true|, \verb|sanitize=false|, \verb|relax=false|, \verb|num_resamplings=1|, \verb|jump_length=1|, and \verb|num_timesteps=1000|. The 50 generated SDF files were concatenated using the shell command \verb|cat|. 

\subsection{GeoLDM}
The code for GeoLDM with commit hash \verb|03ae2031c712a1a6c1678e747bdcdc7a7560e00b| was obtained from the authors’ public code repository \url{https://github.com/MinkaiXu/GeoLDM}. 
The weights for the model trained on GEOM Drugs available in the directory were used. The model was run using the script \verb|eval_analyze.py| with the setting for batch size at 100. 
The model was run 10 times sampling 10,000 molecules each time. The generated TXT files were converted to SDF format and combined into a single file with Open Babel \citep{oboyle2011}.

\subsection{SemlaFlow}
The code for SemlaFlow with commit hash \verb|0c021d663f9feacbfe19e6f3527b2ad98d58ecab| was obtained from the authors' public code repository \url{https://github.com/rssrwn/semla-flow}
The model weights for the model trained on the {GEOM drugs} dataset were obtained via the links in the repository. The model was run once using the `predict` script provided by the authors and the output was a single SDF file. 

\section{Datasets}
\label{sec:datasets}

The Geometric Ensemble Of Molecules (GEOM) was curated by \citet{axelrod2022}. 
The dataset has two subsets: the `QM9' sub set contains 133k molecules with up to 9 heavy (non-hydrogen) atoms and
GEOM's `Drugs' sub set contains 304k drug-like molecules with up to 91 heavy atoms and for all these molecules, at least one conformation annotated with the conformation potential energies is available.
The GEOM Drugs structures used here were obtained from \citet{irwin2024}. 

The DrugBank \citep{knox2024drugbank} contains CA approved, investigational, and withdrawn drugs and their structures. 
Here, DrugBank v5.1.13 released in January 2018 containing 2`185 structures of approved drugs was used. The data set used here is that of the 2'066 drugs which are provided with 3d structures and have the status `approved'.
It is released under a Creative Commons Attribution-NonCommercial 4.0 International License.

\end{document}